\documentclass[aps,prl,twocolumn,citeautoscript]{revtex4-1}
\usepackage{graphicx}
\usepackage{mathrsfs}
\usepackage{amsmath}
\usepackage{amssymb}
\usepackage{amsbsy}
\usepackage{color}
\usepackage[nice]{nicefrac}
\usepackage{float}
\usepackage{lipsum}
\usepackage[a4paper,lmargin=2cm,rmargin=1.2cm,tmargin=2.2cm,bmargin=2.2cm]{geometry}

\usepackage{color}
\definecolor{darkred}{rgb}{0.6,0.,0.}
\definecolor{darkgreen}{rgb}{0.,0.5,0.}
\definecolor{darkblue}{rgb}{0.,0.,0.6}

\usepackage[
breaklinks,
colorlinks=true,
linkcolor=darkred,
citecolor=darkgreen,
urlcolor=darkblue]{hyperref}

\def\be{\begin{eqnarray}}
\def\ee{\end{eqnarray}}

\def\ba{\begin{align}}
\def\ea{\end{align}}

\iftrue

\def\eref{\ref}

\else
\makeatletter
\def\eref{\@ifstar \@refstar \Toref}
\def\Toref #1{\NR@setref {#1}\firstplusthree {#1}}
\makeatother
\def\firstplusthree#1#2#3#4#5{P-#1}

\fi

\def\comment#1{(see comment in source)}

\renewcommand{\-}{\,-\,}

\newcommand{\br}{\mathbf{r}}
\newcommand{\bk}{\mathbf{k}}
\newcommand{\bv}{\mathbf{v}}

\let\oldmarginpar\marginpar
\renewcommand\marginpar[1]{\-\oldmarginpar[\raggedleft\tiny #1]%
{\raggedright\tiny #1}}

\begin{document}

\title{Fractional Chern Insulators in Harper-Hofstadter Bands with Higher Chern Number}

\author{Gunnar M\"oller}
\affiliation{TCM Group, Cavendish Laboratory, University of Cambridge, Cambridge CB3 0HE, United Kingdom}

\author{Nigel R. Cooper}
\affiliation{TCM Group, Cavendish Laboratory, University of Cambridge, Cambridge CB3 0HE, United Kingdom}

\date{August 31, 2015}

\begin{abstract}
The Harper-Hofstadter model provides a fractal spectrum containing topological bands of any integer Chern number, $C$. 
We study the many-body physics that is realized by interacting particles occupying Harper-Hofstadter bands with $|C|>1$. We formulate the predictions of Chern-Simons or composite fermion theory in terms of the filling factor, $\nu$, defined as the ratio of particle density to the number of single-particle states per unit area.  We show that this theory predicts a series of fractional quantum Hall states with filling factors $\nu = r/(r|C| +1)$ for bosons, or $\nu = r/(2r|C| +1)$ for fermions. This series includes a bosonic integer quantum Hall state (bIQHE) in $|C|=2$ bands. We construct specific cases where a single band of the Harper-Hofstadter model is occupied.  For these cases, we provide numerical evidence that several states in this series are realized as incompressible quantum liquids for bosons with contact interactions.
\end{abstract}

\pacs{
73.43.Cd, 	
%
71.10.Pm, 
%
05.30.Pr 	
%
%
03.75.Lm 	
%
%
}

\maketitle

Recently, there has been much progress towards experimental realizations of topological flat bands, such as by light-matter coupling in cold gases \cite{Lin:2009us,Aidelsburger:2013ew, 2013PhRvL.111r5302M, Aidelsburger:2015hm, Jotzu:2014kz, Dalibard:2011gg,Goldman:2014bv} or via spin-orbit coupling in condensed matter systems 
\cite{Tang:2011by, Neupert:2011db, Sun:2011dk}. These systems provide novel avenues for exploring fractional quantum Hall physics 
 in new settings where lattice effects play important roles \cite{Kol:1993wv, Sorensen:2005bt, Palmer:2006km, 2009PhRvL.103j5303M, Kapit:2010ky, Tang:2011by, Neupert:2011db, Sun:2011dk, Sheng:2011ku, Regnault:2011bu, Bergholtz:2013ey, Parameswaran:2013br}.
Furthermore, these ``fractional Chern insulators'' generalize the fractional quantum Hall states of interacting particles in continuum Landau levels to lattice-based systems.

In cases where the underlying topological band has unit Chern number, $C=1$, the states can be continuously connected to conventional fractional quantum Hall states in the continuum Landau level \cite{Scaffidi:2012dx}. However, if the band has Chern number, $C$, of magnitude greater than $1$, no such continuity is possible. The fractional quantum Hall states have features that are particular to the lattice structure. 
The appearance of fractional quantum Hall states for bands with $|C|>1$ has been demonstrated in various lattice models, with unit cells that contain multiple states in the form of distinct sublattices, or in terms of internal degrees of freedom such as spin or color \cite[see, e.g.~][]{Wu:2015id}.
In particular, this has led to the proposal of states at filling factors $\nu = 1/(|C|+1)$ for bosons \cite{2012PhRvB..86t1101W,Liu:2012ek,Sterdyniak:2013du,Wu:2013ii}.

In this Letter we show that this physics of interacting particles in the novel Chern bands can be captured within the Harper-Hofstadter model, in which a magnetic unit cell arises naturally without additional internal degrees of freedom. 
This model leads to a complex energy spectrum as a function of flux $n_\phi=\Phi/\Phi_0$ per plaquette. The low energy bands can have Chern numbers larger than one.  
We show that they can realize the sequences of fractional Chern insulator states for $|C|>1$ discussed for other models, providing
an interpretation of these states  in terms of the composite fermion construction on a lattice \cite{Kol:1993wv,2009PhRvL.103j5303M}. 
Based on these insights, we identify another sequence of fractional Chern insulator states with filling factors $\nu = r/(r|C| +1)$. We show numerical evidence for this sequence from exact diagonalization studies.

The Harper-Hofstadter model has recently been realized, at least for weakly interacting particles, in experiments on ultracold gases \cite{Aidelsburger:2013ew, 2013PhRvL.111r5302M, Aidelsburger:2015hm}. Further realizations have also been obtained for a triangular moir\'e lattice in graphene flakes deposited on boron nitride \cite{Dean:2013bv, Ponomarenko:2013hl, Hunt:2013ef}. 
Our results demonstrate that, under suitable conditions (particle density, flux density and temperature), these systems have the possibility to explore a wide range of the novel physics of fractional Chern insulators.

The rich physics of charged particles in a two-dimensional plane subjected to a perpendicular magnetic field $B$ and a periodic potential, or physically equivalent models emulating this scenario, arises from its two competing length scales: the magnetic length $\ell_0=\sqrt{\hbar/eB}$ and the lattice scale $a$. 
Hence, the problem brings to play the commensurability of these two scales, as first analyzed by Harper \cite{Harper:1955bj}. The resulting fractal structure of the single-particle spectrum was revealed by Azbel \cite{Azbel:1964tk} and the full spectrum solved numerically and characterized by Hofstadter \cite{Hofstadter:1976wt}.
As shown by Wannier \cite{Wannier:1978eb}, the Azbel-Hofstadter recursion relations imply that the total number of states per unit area $n_s$ below each gap varies linearly with the flux density $n_\phi$, and further that their relationship is described by the Diophantine equation
\be
\label{eq:diophantineN}
n_s = C n_\phi - D, \quad C,D \in \mathbb{Z} \,.
\ee
The work by Streda \cite{Streda:1982dn}, as well as Thouless et al.~\cite{Thouless:1982kq, Niu:1985ui} explains the physical relevance of Wannier's result. For noninteracting fermions, one expects incompressible states at density $n=n_s$, with Hall conductivity given by Streda's formula as
\be
\sigma_{xy} = \frac{e}{\Phi_0} \frac{\partial n}{\partial n_\phi} = C \frac{e^2}{h}.
\ee
Thouless et al. derive the Hall conductivity by direct calculation from a Kubo formula, obtaining the integer quantization from the topological nature of the resulting expression, namely that $C=\sum \mathcal{C}_i$ where $\mathcal{C}_i$ is the Chern number of the $i$-th occupied band \cite{Thouless:1982kq}. Thus, the integer $C$ in Eq.(\ref{eq:diophantineN}) is seen to be the net Chern number of the bands contributing to the $n_s$ states below the energy gap.
Solutions to (\ref{eq:diophantineN}) exist for any $C$ in which $n_s$ arises from a single band (see below), 
establishing the presence of bands of any Chern number in the Hofstadter spectrum (albeit with rapidly decreasing gaps for large $\mathcal{C}$) \cite{Osadchy:2001jm,Avron:2003ui}. 

To realize the nontrivial Chern bands of the Harper-Hofstadter model, it is sufficient to create a tight-binding lattice system with complex hopping elements between nearest neighbor sites, a fact exploited for realizations of the model in cold gases \cite{Dalibard:2011gg,Aidelsburger:2013ew, 2013PhRvL.111r5302M}. From the analysis of the magnetic translation group \cite{Harper:1955ft, 1964PhRv..136..776Z}, it follows that at a rational flux density $n_\phi=p/q$, the Harper-Hofstadter Hamiltonian admits a periodic representation on a magnetic unit cell (MUC) comprising $q$ sites, i.e., an area enclosing an integer number of flux quanta and an integer number of plaquettes of the lattice.
The single-particle Hamiltonian then takes the tight-binding form
\be
\label{eq:Hamiltonian}
\mathcal{H}_\text{sp} = - \sum_{i,j} t_{ij} e^{\phi_{ij}} \hat a_j^\dagger \hat a_i + h.c.,
\ee
in which the phases $\phi_{ij}$ are invariant under translations of the MUC. 
(The choice of the MUC and the vector potential $\mathbf{A}$ make up the remaining space of gauge choices.)
One can therefore consider the $q$ sites of the MUC as sublattices $\alpha=1,\ldots,q$ of a general $q$-site tight-binding model and solve via Bloch's theorem.
Note that the origin of the MUC can be chosen on any site of the lattice, so the problem has an additional $q$-fold symmetry. 
We give an analysis of the single-particle properties in the supplementary material \cite{supp-mat}, taking care to respect this symmetry \cite{Jackson:2014vj}. 

To search for strongly-correlated phases, it is useful to identify situations in which there is a 
manifold of low-energy single-particle states (one band, or several closely spaced bands) that is well separated from higher-energy bands. For now, we focus on the case in which this manifold is a single band, with a large gap to the next band.
The largest gap in the Harper-Hofstadter spectrum corresponds to the lowest Landau level, with $C=1$, $D=0$. Here, we seek more general states, and consider the next-largest gaps found at the first level of the Hofstadter hierarchy. These appear  in the vicinity of cell boundaries close to the simple rational flux densities $n_\phi=1/Q$, ($Q>1$), where the energy bands become exponentially flat in terms of both their energy dispersion and their Berry curvature. Hence, these bands are well suited to support incompressible fractional quantum Hall states \cite{Palmer:2006km,Palmer:2008cq,2009PhRvL.103j5303M,2012PhRvL.108y6809H}. The gaps near points $(n_s,n_\phi)=(0,Q^{-1})$ are described by the Diophantine equation (\ref{eq:diophantineN}) with $C\equiv s Q$, and $D= \text{sgn}(C) = s$, with $s=\pm 1$ for the bands at $n_\phi \gtrless 1/Q$. We are particularly interested in cases where we find a single band below this gap, i.e., where $n_s(n_\phi=p/q)=1/q$, and thus $q=Qp - s$, with corresponding flux densities
\be
\label{eq:SingleBandCases}
n_\phi = \frac{p}{|C|p - \text{sgn}(C)},\quad p \in \mathbb{N}.
\ee
Below, we take $p> 2$ to ensure that the band belongs to the subcell nearest $n_\phi=1/|C|$. We will also consider higher band gaps at the same flux densities, which can be seen as fractal replicas of the $r$-th continuum Landau level, for which
\be
\label{eq:diophantineEqRthSubcell}
 n_s = r(C n_\phi-\text{sgn}C), \quad r\in \mathbb{Z}\backslash \{0\},
\ee
and where $rC\gtrless 0$ for the flux densities $n_\phi \gtrless 1/Q$.

We now discuss the many-body physics of interacting particles in the Harper-Hofstadter model, described by the Hamiltonian
\be
\label{eq:FullHamiltonian}
\mathcal{H} = \mathcal{H}_\text{sp} + \frac{1}{2}\sum_{i,j} V(\br_i - \br_j) :\hat n_i \hat n_j:,
\ee
with site labels $i$, $j$, and $:\hat n:$ denoting normal ordering of the density operators. Let us first review the predictions of Chern-Simons theory \cite{Kol:1993wv}, or equivalently the lattice composite fermion picture \cite{2009PhRvL.103j5303M}, and translate these results into the language used for the analysis of Chern insulators. 
The basic premise of this approach is that the interaction includes a sufficiently strong short-range repulsion in order to favor ``flux attachment,'' which keeps the particles at a distance from each other, thus minimizing interaction energy.
The composite fermion Ansatz translates this idea into a trial wave function of the form $\Psi_\text{trial}(\br_1,\ldots,\br_N) = \mathcal{P}_\text{LEM} \Psi_\text{J} (\{\br_i\})
\times \Psi_\text{CF}  (\{\br_i\}),$
where both the Jastrow factor $\Psi_\text{J}$ and the composite fermion wave function $\Psi_\text{CF}$ vanish when the positions of two particles coincide, and $\mathcal{P}_\text{LEM}$ denotes the projection onto the relevant low-energy manifold of single-particle states. 
For the case of bosons (fermions), one needs to attach an odd (even) number $k$ of flux quanta to the particles, so as to obtain an effective problem of weakly interacting composite fermions (CF) experiencing an effective flux density $n_\phi^*$ relating to the externally applied flux via
\be
\label{eq:flux_attachment}
n_\phi =  k n + n_\phi^*,  \quad k\in \mathbb{Z}\backslash \{0\}.
\ee
If the CFs behave as weakly interacting particles, they will form incompressible (topological) insulating states when filling an integer number of bands. Their band structure is given by a Harper-Hofstadter Hamiltonian with flux density $n_\phi^*$. 
The densities $n_s^*$ at which filled bands are realized are therefore given by a Diophantine equation of the form (\ref{eq:diophantineN}) for the composite fermion system, $n_s^* = C^* n_\phi^* - D^*$, with integer parameters $C^*$, $D^*$. In composite fermion theory, one can explain the fractional QHE as an integer QHE of composite fermions \cite{Jain:1989tq}, and for the lattice case one thus predicts incompressible states at $n=n_s^*$. Hence, with (\ref{eq:flux_attachment}), we find
\be
\label{eq:CFDensity}
n = \frac{C^* n_\phi - D^*}{kC^* + 1} > 0.
\ee
Choices of the parameters $C^*$ and $D^*$ for the composite fermion gap yield various candidates for incompressible states in the spectrum, given in terms of density, as illustrated previously as Fig.~1 in Ref.~\onlinecite{2009PhRvL.103j5303M}. 

In order to relate the densities (\ref{eq:CFDensity}) to the ``filling
factor'' of FQH systems there are several choices that can be
made. One choice is to consider the ratio, $\nu_\phi \equiv n/n_\phi$,
of particle density to flux density, which is natural in the continuum
limit, $n_\phi\to 0$, where the bands of the Harper-Hofstadter model
reduce to continuum Landau levels.  More generally, and to allow
connections to fractional Chern insulator models, the natural filling
factor to consider is the ratio $\nu \equiv n/n_s$ of the particle
density to the number density of single-particle states in the
low-energy manifold (e.g. the lowest energy band if this manifold is a
single band).
We replace $n_\phi$ in Eq.~(\ref{eq:CFDensity}) via (\ref{eq:diophantineN}) and obtain 
\be
\label{eq:CFRelationOfNandNs}
n\left(\frac{1+k C^*}{C^*}\right) + \frac{D^*}{C^*}  = \frac{n_s}{C} + \frac{D}{C}.
\ee
In general, all parameters $\{k, C, C^*, D, D^*\}$ contribute to determine eligible states. However, as we now describe, for some important  cases the ratios
 $\frac{D^*}{C^*}  = \frac{D}{C}$ are equal, and the states can be characterized by a fixed filling factor $\nu=n/n_s$.

It is instructive to consider special cases. Firstly, the fractional quantum Hall states in a $C=1$ band are recovered by choosing the manifold as the (lattice equivalent of the) lowest Landau level ($C=1$, $D=0$), and taking general integers $C^* = t$, $D^*=0$ for filling composite fermion states in the $t$-th Landau level. One recovers the usual Jain series of states $\nu=t/(kt+1)$. (In this case, $n_s =n_\phi$, so the two definitions of filling factor coincide.)

Secondly, we can take both the CF bands and the effective low-energy bands in the same subcell of the Hofstadter spectrum, close to the flux $n_\phi = 1/|C|$. As a concrete example, consider the gaps (\ref{eq:diophantineEqRthSubcell}) and choose to fill $r$ bands of composite fermions such that $C^* = r C$, and $D^*=r \, \text{sgn}(C)$, and choose the lowest band with the given $C$, and $D=\text{sgn}(C)$ for the manifold of single-particle states. 
We obtain states with filling factors \footnote{Note, the positivity of (\ref{eq:CFDensity}) implies $kC>0$, which fixes the required sign of $k$.}
\be
\label{eq:FillingFactorJainEquiv}
\nu^{C^*=rC} = \frac{n}{n_s} =  \frac{r}{ r |k C| + 1}, \quad r \in \mathbb{Z}\backslash \{0\},
\ee
where states with $r<0$ represent the generalization of negative flux attachment.
In general, the low-energy manifold supporting these states will have many bands, but in the cases (\ref{eq:SingleBandCases}) this reduces to a single band.

The sequence of filling factors (\ref{eq:FillingFactorJainEquiv}), valid for any Chern number $C\neq 0$, is a core result of our Letter \footnote{Note that the physics in time-reversal symmetric bands with Chern number $C=0$ yields very different phenomena, such as Bose condensates \cite{2010PhRvA..82f3625M} or supersolids \cite{2012PhRvL.108d5306M}.}.

Several remarks are in order. The case with $r=1$ can be seen as an analogue of the Laughlin state, in the sense that a single band of composite fermion states is filled.
From the previous studies of the Laughlin state on the lattice in a $C=1$ band, we can infer useful intuition on the likely stability of such states. 
The Laughlin state was shown to be stable up to flux densities $n_\phi\simeq 0.4$, i.e. it persists through 80\% of the region in which a gap is open \cite{Hafezi:2007gz}. Likewise, the $\nu=2/3$ hierarchy state was seen to be stable up to $n_\phi\simeq 0.3$ \cite{2009PhRvL.103j5303M}.
In the case of the states stabilized in subcells with $|C|>1$ bands, we note that the bands tend to have less dispersion, albeit maybe larger fluctuations in the band geometry. By analogy, the family of states (\ref{eq:FillingFactorJainEquiv}) can be expected to be stable at a substantial distance from the respective cell boundaries.

The reader will note that the prediction of composite fermion theory (\ref{eq:FillingFactorJainEquiv}) includes the Abelian states at filling factors $\nu=1/(|kC|+1)$ that have recently been described in studies of Chern bands with Chern number $|C|>1$ \cite{2012PhRvB..86t1101W,Liu:2012ek,Sterdyniak:2013du}, for both bosons and fermions, and which were described in terms of $C$ flavour states \cite{Palmer:2006km, Palmer:2008cq, 2012PhRvL.108y6809H, Wu:2013ii}. The derivation presented here demonstrates that these states are predicted also by the concept of flux attachment \cite{Kol:1993wv} leading to CF wave functions of the form described in Ref.~\cite{2009PhRvL.103j5303M}.
While the $C$-flavour/multilayer language appears to require $C$ copies of a $C=1$ Brillouin zone, implying finite size geometries with a number of states $N_s \mod C =0$, it was shown that a color-entangled formulation remedies this constraint \cite{Wu:2013ii}. Note that the hierarchy wave functions following from the CF construction \cite{2009PhRvL.103j5303M} similarly do not require any constraint on the lattice geometry. 

The composite fermion theory makes a more general prediction, in that it does not require that the single particle states making up the manifold $n_s$ are from a single band, as $n_\phi$ can vary continuously in (\ref{eq:CFRelationOfNandNs}). Indeed, perturbing a stable quantum liquid formed in a single-band configuration by an infinitesimal change in $n_\phi$, the low-energy manifold splits up into (possibly infinitely) many bands, but we expect that the physics of the phase should be robust under this perturbation, providing a notion of adiabatic continuity that allows us to connect any band to the limit of the perfectly flat general Chern bands obtained as $n_\phi\to 1/C$ \cite{supp-mat}, in line with the behaviour seen in $C=1$ Harper-Hofstadter bands \cite{Bauer:2015vp}.

The filling factors (\ref{eq:FillingFactorJainEquiv}) are analogous to the hierarchy states, which have been observed in Chern number one fractional Chern insulators \cite{2009PhRvL.103j5303M,Lauchli:2013ck,2013PhRvB..87t5136L}. Their properties, in terms of quasiparticle charges and statistics were predicted by Kol and Read \cite{Kol:1993wv}, as summarized in \cite{supp-mat}.
Unlike the lowest Landau level, $|C|>1$ bands support states with negative flux attachment ($r<0$) even for $|k|=1$, $r=-1$, so the corresponding series of states leads to novel filling factors. Numerical evidence for the $\nu=1$ state in a $C=2$ band obtained for $Ckr=-2$ was provided by the current authors in Ref.~\cite{2009PhRvL.103j5303M}, which is a special case in that it realizes an integer quantum Hall effect of bosons \footnote{We thank B. B\'eri for pointing out a possible relation to the boson integer quantum Hall effect (bIQHE) \cite{Senthil:2013kt}}.

The limit of filling many CF Landau levels, $r\to \infty$, which represents the equivalent of the half-filled Landau level, converges to
\be
\label{eq:LimitingFilling}
\lim_{r\to\infty} \nu^{C^*=rC} = \frac{1}{|kC|}.
\ee
At these points, the composite fermion spectrum resembles a Fermi-sea, as the band-gaps between the composite fermion levels decrease as $1/r$ and evolve into a quasi-continuum. In analogy to the half-filled continuum Landau levels, one may ask whether this filling can be susceptible to the equivalent of a CF pairing instability, or possibly more exotic states. In the $C=1$ case, the possibility of a Moore-Read state at $\nu=1$ is well known \cite{Sterdyniak:2012jo}.
For the $C=2$ band near $n_\phi=1/2$, a paired phase has  been described in a related continuum model \cite{2012PhRvL.108y6809H, Moller:2014fa}, though the model does not provide a quantitative description of $C=2$ bands \cite{Harper:2014fq}. If realized, it is expected that a paired phase at the fillings (\ref{eq:LimitingFilling}) would be non-Abelian in Chern bands with $|C|$ \emph{odd}, while Majorana quasiparticles will likely pair up for even Chern bands and thus recombine to yield an Abelian phase \cite{Moller:2014fa}.

\begin{figure}[t]
\begin{center}
\includegraphics[width=0.95\columnwidth]{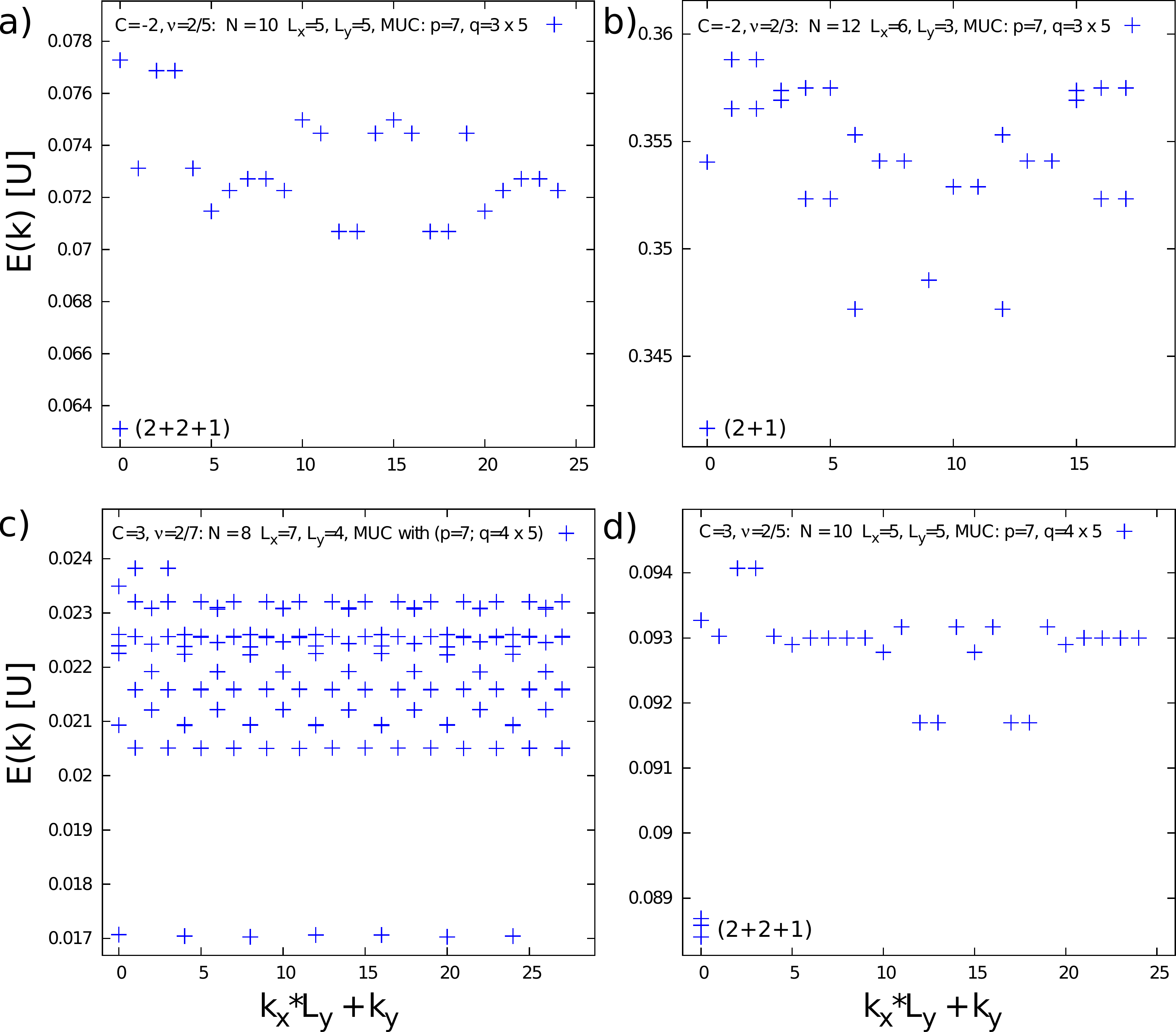}
\caption{Spectra for Chern insulators of the series (\ref{eq:FillingFactorJainEquiv}) in Harper-Hofstadter bands at flux densities from (\ref{eq:SingleBandCases}). For Chern number $|C|=2$ we choose flux density $n_\phi = 7/15$, showing (a) $\nu=2/5$ for $N=10$ (b) $\nu=2/3$ for $N=12$ bosons. For $C=3$ we take $n_\phi = 7/20$, and show (c) $\nu=2/7$ for $N=8$, (d) $\nu=2/5$ for $N=10$. For lattice geometries, see legend. Multiplet structure of ground states shown in parentheses.}
\label{fig:Spectra}
\end{center}
\end{figure}

A case that has not yet been explored is the Abelian series of states (\ref{eq:FillingFactorJainEquiv}).
We examine the evidence for the presence of these composite fermion or hierarchy states 
on the basis of the band-projected Hamiltonian within the low-energy manifold, focusing on the single-band cases (\ref{eq:SingleBandCases}). The corresponding Hamiltonian, $\mathcal{H}_\text{proj} = \mathcal{P}_\text{LEM} \mathcal{H} \mathcal{P}_\text{LEM}$,
can be studied in the same framework as other fractional Chern insulator models \cite{Regnault:2011bu}. 
The residual dispersion of bands in the low-energy manifold could be of interest for studying phase transitions between fractional quantum Hall liquids and condensed phases of bosons or Fermi liquid like states of fermions, respectively. 
However, here we choose to neglect the residual band dispersion, particularly as it vanishes quickly as $n_\phi\to 1/|C|$ \cite{supp-mat}. Furthermore, we focus on the case of bosons with contact interactions $V_{ij}=U \delta_{ij}$.

Our numerical study shows evidence supporting the existence of gapped quantum liquids at several filling factors of the series (\ref{eq:FillingFactorJainEquiv}). 
Firstly, states are found for the cases $r=1$, $|k|=1$, where the predictions of the filling factor $\nu=1/(|C|+1)$ coincide between the composite fermion theory and the analyses in terms of Halperin multi-component \cite{Palmer:2006km, Palmer:2008cq, 2012PhRvL.108y6809H} or color-entangled states \cite{Wu:2013ii}. 
The integer bosonic quantum Hall state with $r=-1$ was discussed in Ref.~\cite{2009PhRvL.103j5303M}. Here, we present evidence for additional states, such as those with $|r|=2$, with two filled composite fermion bands. 
In figure \ref{fig:Spectra}, we show their spectra for Chern bands with $|C|=2$ [panels a), b)], and $C=|3|$ [panels c), d)]. 
All cases show the correct ground state degeneracies predicted by CF theory ($d_\text{GS}=|1+kC^*|$, see Refs.~\cite{Kol:1993wv, supp-mat}). 
The states with positive flux attachment, $\nu=2/5$ ($|C|=2$) and $\nu=2/7$ ($|C|=3$) have the clearest signature in terms of the magnitude of the gap to the average state spacing of excitations. 
The states with negative flux attachment, $\nu=2/3$ ($|C|=2$) and $\nu=2/5$ ($|C|=3$) also show a distinct separation of energy scales. 

\begin{figure}[t]
\begin{center}
\includegraphics[width=0.95\columnwidth]{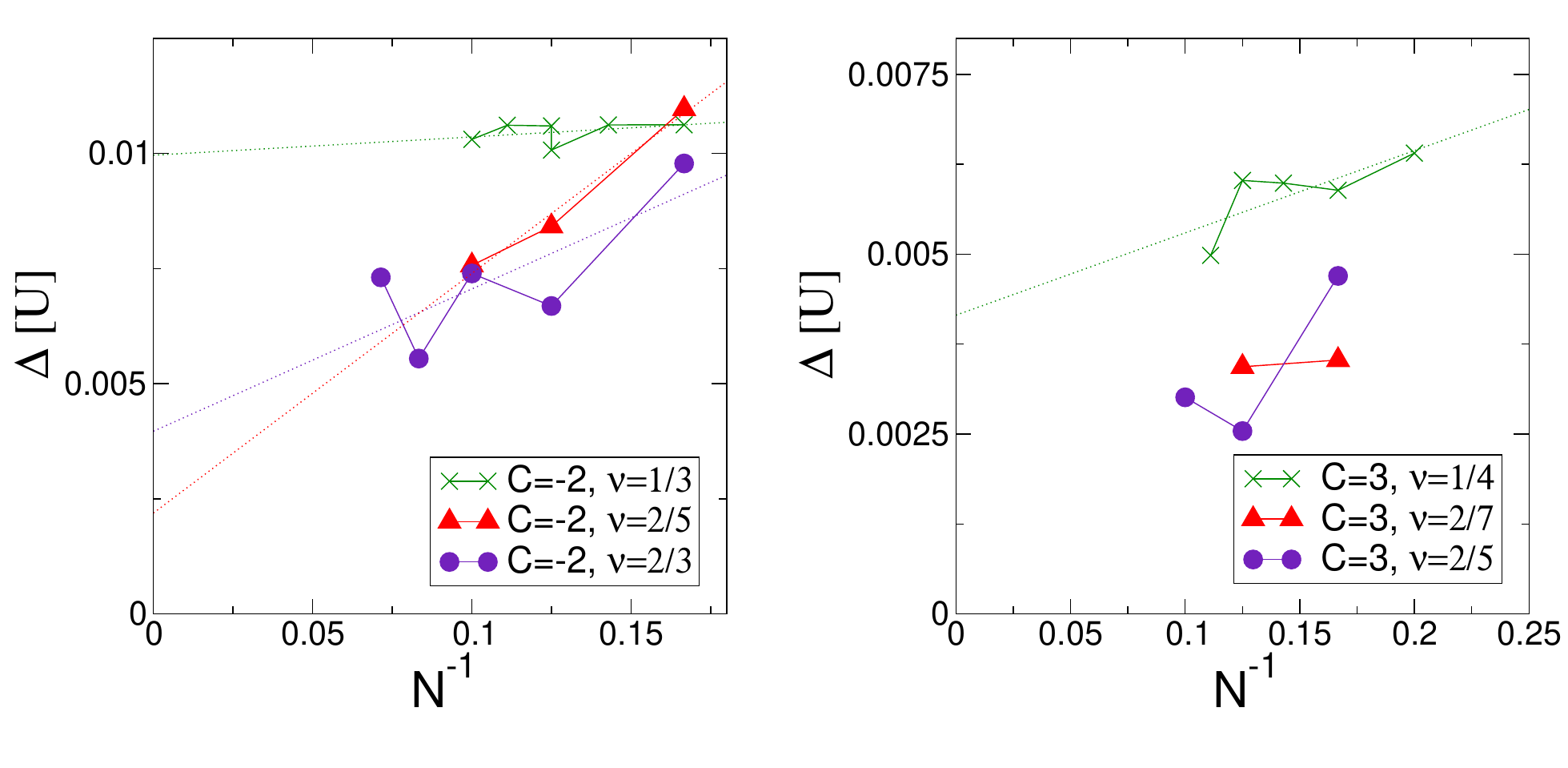}
\caption{Finite size scaling of the gap for states of the series (\ref{eq:FillingFactorJainEquiv}), with $r=+1,+2, -2$. Left: a $C=-2$ band ($n_\phi=7/15$) for states at $\nu=1/3$, $\nu=2/5$, and $\nu=2/3$. Right: Finite size scaling for $C=3$ states ($n_\phi=7/20$) at $\nu=1/4$, $\nu=2/7$, and $\nu=2/5$. 
}
\label{fig:GapScaling}
\end{center}
\end{figure}

A finite size scaling of the gap gives us further indications of the stability of these phases. 
Figure \ref{fig:GapScaling} shows the gap scaling for several filling factors in the same $|C|=2$ and $|C|=3$ bands as above. 
All systems we examined show the expected ground state degeneracy. In both bands, the largest gap is found for the $r=1$ state of the series (\ref{eq:FillingFactorJainEquiv}), while the $r=\pm 2$ states have a slightly smaller gap in the finite size systems. 
The extrapolation to the thermodynamic limit is consistent with a non-zero gap for both the $r=1$ and $r=\pm 2$ states for $|C|=2$. Data for the $C=3$ cases are both noisier and include fewer system sizes. Nonetheless, the results are consistent with a non-zero gap. 

In the supplementary material, we briefly discuss particle entanglement spectra of our states, as well as spectral flow under flux insertion \cite{supp-mat}. In addition to these ground state properties, we have examined spectra under addition of ``flux,'' i.e., under changes of system size at fixed $N$. We find low-lying bands of states consistent with an interpretation as quasiparticle states of an underlying quantum Hall liquid. We leave the detailed analysis of these features for a future publication.

In conclusion, we have translated the composite-fermion or Chern-Simons theory into the language of fractional Chern insulators, leading to the prediction of a series of states with filling factors $\nu=r/(r|kC|+1)$, for bosons ($|k|=1$) or fermions ($|k|=2$). This includes, and provides an alternative description for, the series of states $\nu=1/(|kC|+1)$ that were observed in the literature on FCI for $|C|>1$. 
We have identified flux densities where a single isolated band of Chern number $C$ occurs at the bottom of the Hofstadter spectrum. Finally, we have studied the many-body states of bosons with contact interactions under the projection into these Chern bands, identifying gapped states with the ground state degeneracies predicted by theory. 
While previous evidence had been given for the bosonic integer Chern insulator state with $\nu=1$, $r=-1$, $|C|=2$ \cite{2009PhRvL.103j5303M}, which was obtained for a hard-core interaction and without applying a band projection, the current results provide evidence for the wider applicability of composite fermion theory, and its validity also for the band-projected Hamiltonian. Our results can be extended to general $|C|>1$ bands in other tight-binding models, and to the effective continuum limit $n_\phi\to 1/|C|$ via principles of adiabatic continuity.
Further investigations should focus on the stability of fermionic states, the role of long-range interactions and the detailed analysis of the ground states and excitations in terms of the composite fermion trial wave functions.

\begin{acknowledgments}
We acknowledge useful discussions with Antoine Sterdyniak, Nicolas Regnault, Steve Simon,  Zohar Ringel, Benjamin B\'eri, and Thomas Scaffidi. We also thank Rahul Roy for discussions and for sharing Ref.~\cite{Bauer:2015vp} in advance of publication.
The authors acknowledge support from the Leverhulme Trust under Grant No.~ECF-2011-565, from the Isaac Newton Trust, and by the Royal Society under Grant No.~UF120157 (G.M.),
as well as by Engineering and Physical Sciences Research Council Grant Nos. EP/J017639/1 and EP/K030094/1 (N.R.C.).
\end{acknowledgments}

%

 \vskip3em

\begin{center}
\textbf{\large Supplemental Material}
\end{center}

\setcounter{secnumdepth}{2}

\renewcommand{\thesection}{\Alph{section}}
\numberwithin{equation}{section}

\renewcommand{\thefigure}{S-\arabic{figure}}

\section{Single Particle Harper-Hofstadter Hamiltonian}
\label{sec:SingleParticle}

As stated in the main text, the Harper-Hofstadter Hamiltonian can be written in a gauge with explicit periodicity given by multiples of the lattice vectors $\{\mathbf{v}_1, \mathbf{v}_2\}$ of an underlying periodic potential. To this end, we require a pair of magnetic translation operators $T_M$ \cite{app-1964PhRv..134.1602Z} that respect the translations of the lattice, i.e., we seek
\ba
t_1=T_M(l_1 \mathbf{v}_1),\quad t_2=T_M(l_2 \mathbf{v}_2),
\end{align}
and we choose $l_i$ such that the area swept by the translations encloses the smallest possible integer number of flux quanta $N_\phi$. The cell can take geometries satisfying
\be
\label{eq:contraintForMUC}
l_1l_2 |\mathbf{v}_1 \wedge \mathbf{v}_2| = N_\phi^c \ell_0^2,
\ee 
where $l_1l_2=q$ and $N_\phi^c=p$ yield the target flux density of $n_\phi=p/q$.
Solutions to (\ref{eq:contraintForMUC}) provide a ``magnetic unit cell'' (MUC) for the Hofstadter problem.

The phases $\phi_{ij}$ in the single-particle Hamiltonian (\eref{eq:Hamiltonian})
\cite{app-Note1}
for a given gauge take the explicit form
\be
\phi_{ij} = \frac{e}{\hbar} \int_{\br_i}^{\br_j} \mathbf{A}\cdot d\mathbf{l} + \delta\phi_{ij},
\ee
where $\delta\phi_{ij}$ encodes the additional phase factor generated by the magnetic translation operator (relative to canonical translations) for any hopping terms crossing the boundary of a MUC. For a given MUC with translations $\{l_i \bv_i\}$, the entries $t_{ij}\exp(i \phi_{ij})$ of the single-particle tight-binding Hamiltonian are periodic under canonical translations of the MUC. Hence, the many-body Hamiltonian can be simulated on finite-size tori of size $N_\text{cell} = L_1 \times L_2$ MUCs. For the Hofstadter model in the square geometry, we also denote indices $i$ for $l_i, L_i$ by the corresponding coordinate directions $x$, $y$. 

For our studies of the many-body Hamiltonian under band projection, 
we focus on the single-band cases (\eref{eq:SingleBandCases}) and retain $N_s=N_\text{cell} = L_x \times L_y$ momentum sectors. Note that the total number of sites $N_\text{site}=N_x \times N_y \equiv L_x l_x \times L_y l_y \gg N_s$, allowing for a large reduction of the corresponding Hilbert-space. The total particle density $n_s$ is small, hence band projection is expected to be valid even for strong repulsive interactions. Indeed, the states of the $r=-1$ series were found to be stable for hard-core interactions in Ref.~\cite{app-2009PhRvL.103j5303M}.

\begin{figure*}[t]
\begin{center}
\includegraphics[width=0.99\textwidth]{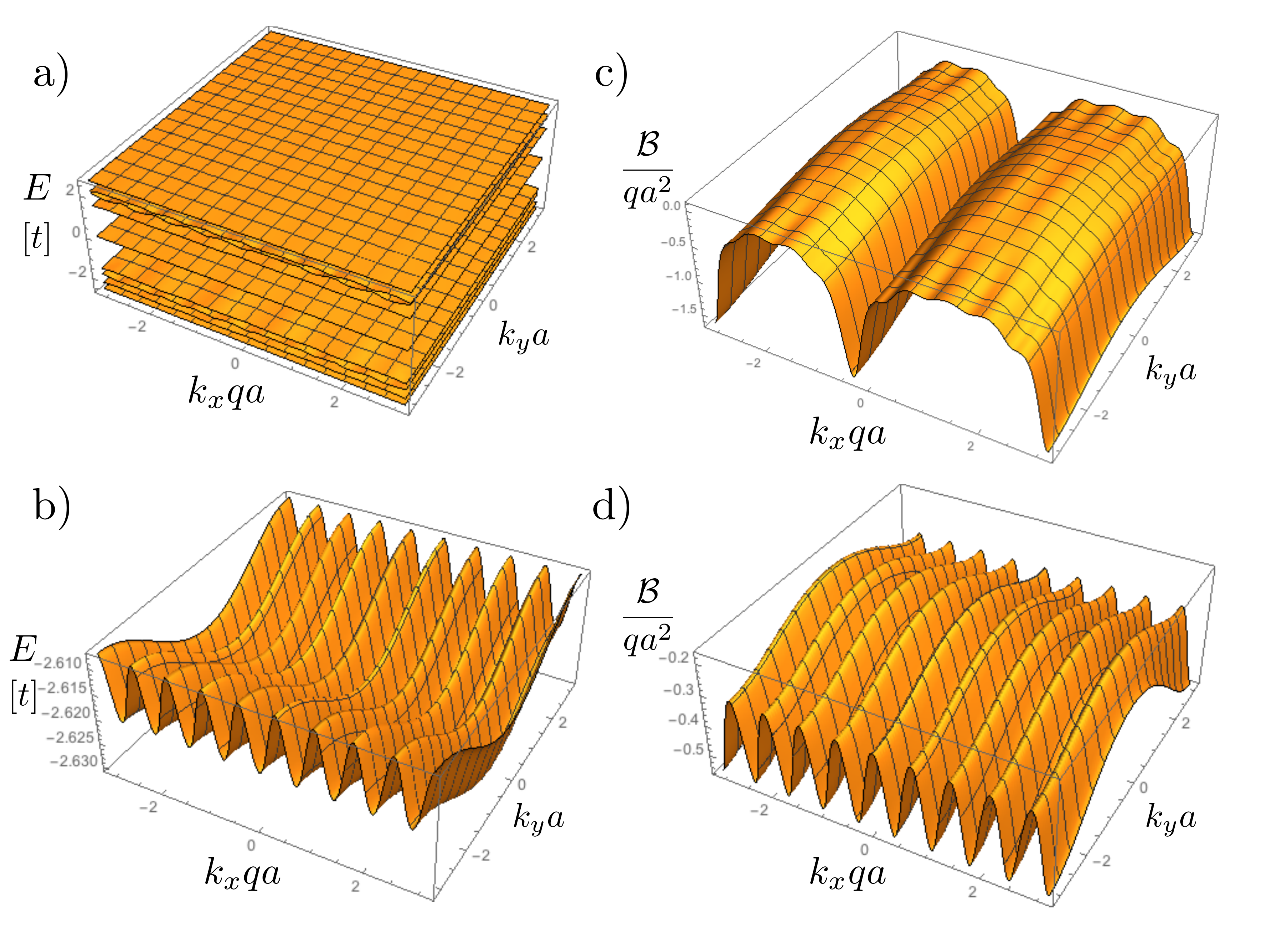}
\caption{Single-particle properties for the Hofstadter Hamiltonian at flux-density $n_\phi=4/9$ for a $l_x \times l_y = 9\times 1$ magnetic unit cell ($q=9$): a) Dispersion for all nine bands, b) Detailed view for the dispersion of the lowest band, c) Berry curvature of the lowest band $\mathcal{B}_\text{UC}$, when Fourier transforms are taken relative to the unit cell (incorrect, see main text), d) canonical Berry curvature of the lowest band $\mathcal{B}$.
}
\label{fig:HofstadterSingleParticle}
\end{center}
\end{figure*}

\section{Properties of Harper-Hofstadter Bands}
\label{sec:FlatBands}

The Harper-Hofstadter bands are generally weakly dispersive, particularly at the extremities of the spectrum. We display an example spectrum in Fig.~\ref{fig:HofstadterSingleParticle}, for the case of $n_\phi=4/9$ given in Harper's gauge of an $l_x \times l_y = 9\times 1$ magnetic unit cell. On the scale of the entire spectrum, the dispersion is negligible (a), while detailed examination reveals a residual dispersion, which carries a $q$-fold symmetry along the extended direction of the chosen magnetic unit cell (b). Owing to this high symmetry, the Hofstadter spectrum is well suited to illustrate a subtlety in calculating the Berry curvatures \cite{app-Jackson:2014vj}, two versions of which are shown in panels c) and d) for the lowest band.
In the former case, we take Fourier-transforms with respect to the position of the unit cell, defining Bloch functions in reciprocal space via $\tilde u_\alpha(\bk) = N_c^{-1/2} \sum_{\mathbf{R}} e^{i \bk \mathbf{R}} u(\mathbf{R} + \rho_\alpha)$, with sublattices labelled by $\alpha$ at sublattice offsets $\rho_\alpha$ and origins of MUCs denoted by $\mathbf{R}$. From these Bloch states, we obtain the first of the Berry curvatures, $\mathcal{B}_\text{UC}$, shown in panel c), which breaks the (translational) symmetry of the problem. Taking the canonical Fourier transform with respect to the site position, with 
$u_\alpha(\bk) = N_c^{-1/2} \sum_{\mathbf{R}} e^{i \bk (\mathbf{R} + \rho_\alpha)} u(\mathbf{R} + \rho_\alpha)$, we obtain the Berry curvature $\mathcal{B}$ shown in panel d), which manifestly respects the translational symmetry of the problem. These findings illustrate the general view that the position of sublattices in the unit cell must be taken into account in calculating the Berry curvature, as physical properties probing $\mathcal{B}$ derive from the canonical position operator \cite{app-Jackson:2014vj}.

\begin{figure}[t]
\begin{center}
\includegraphics[width=0.99\columnwidth]{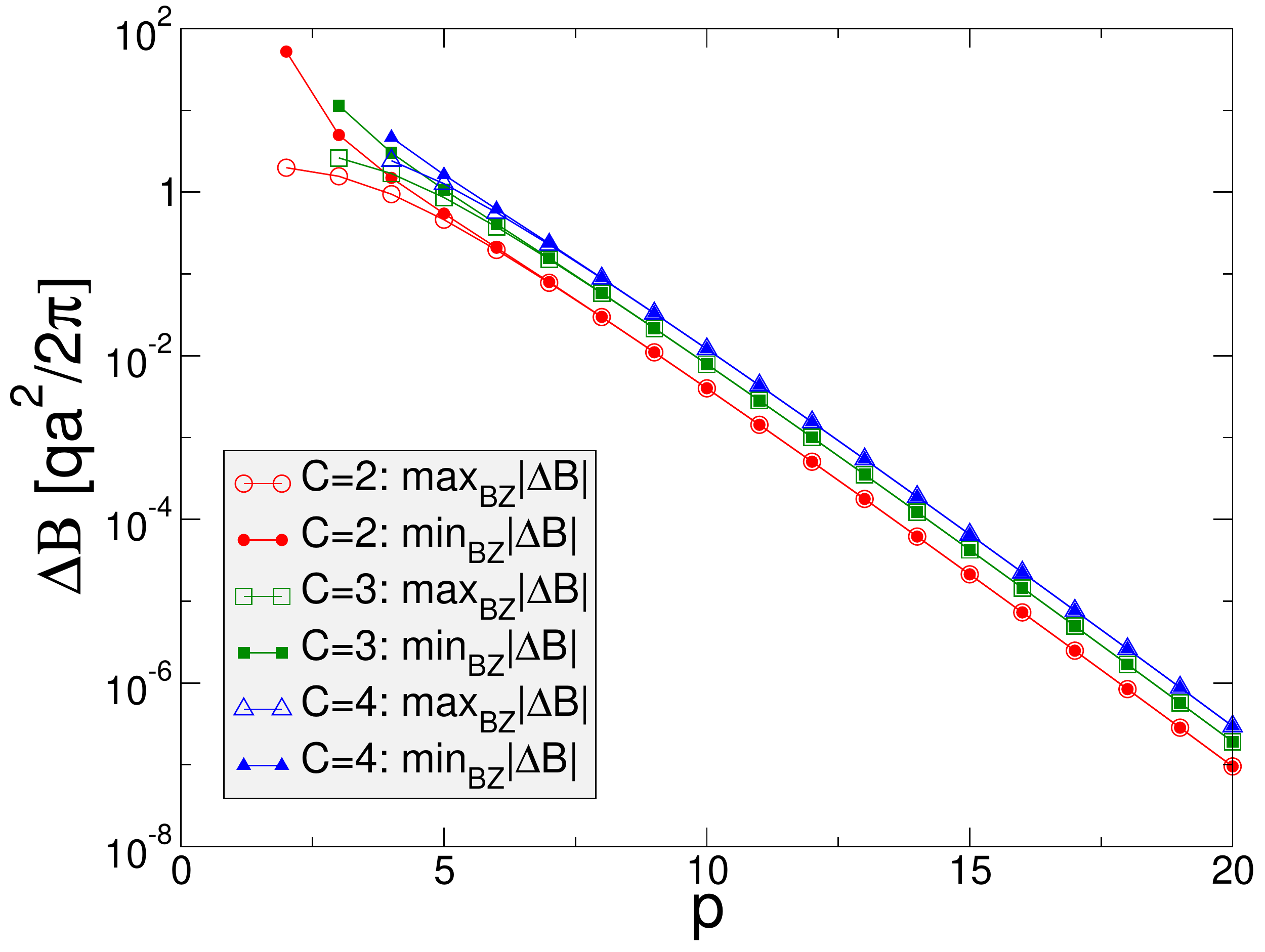}
\caption{Deviation of the canonical Berry curvature from its average value, defined as $\Delta B = |\mathcal{B} - Cq/(2\pi)|$ at the maxima and minima of $\mathcal{B}$ within the Brillouin zone (in units of $q/2\pi$). 
Data is given for the lowest bands the flux densities according to the series of Eq.~(\ref{eq:SingleBandCases}), with $n_\phi=p/q=p/(|C|p-\text{sgn}C)$. Note the logarithmic scale, so the decay is exponential as $n_\phi\to 1/|C|$.
}
\label{fig:HofstadterBerryFlatness}
\end{center}
\end{figure}

In Fig.~\ref{fig:HofstadterBerryFlatness}, we demonstrate how the Berry curvature of the Harper-Hofstadter bands becomes perfectly flat near the points $n_\phi\to 1/|C|$. We show that the extremal values of the Berry curvature approach the expected average value exponentially in the parameter $p$ of Eq.~(\eref{eq:diophantineEqRthSubcell}) in the main text. A similar exponential behavior is seen in the band dispersion. See also Ref.~\cite{app-Harper:2014fq} for a discussion.

\section{Adiabatic continuity of quantum liquids in the Hofstadter spectrum}
\label{sec:adiabatic_continuity}

There are two different kinds of adiabatic continuity which are relevant in the context of the Harper-Hofstadter model.

Firstly, we note that the composite fermion theory incorporates an inherent notion of continuity for both $n$ and $n_\phi$ in the definition of preferred densities, and thus of the many-body spectrum under changes of these parameters satisfying (\ref{eq:CFRelationOfNandNs}). The prediction is that a system in which the low-energy manifold consists of a single band should manifest the same topological order as a system nearby in parameter space, in which $n_\phi$ is an irrational number, and this manifold fragments into infinitely many bands. In this sense, band filling is not the most useful way to analyze the problem. Rather, the useful number of states $n_s$ to consider is contained within the (ensemble of) band(s) below a distinct gap in the Hofstadter spectrum. In particular, this sense of adiabatic continuity allows us to change the flux density gradually until $\Delta n_\phi \equiv n_\phi - 1/|C|$ is small, and the effective magnetic length becomes large. This yields an effective continuum limit, providing the  equivalent of a Landau-level with Chern number $C$, in which both the band dispersion and the band geometry are perfectly flat. 

Secondly, we can use the predictions of composite fermion theory to infer the behavior of general single-particle bands with Chern numbers $|C|>1$. 
In most models of Chern bands, we fix the number of sites per unit cell, and hence the number of bands as the elementary input for the problem. Focusing on the cases where the low-energy manifold consists of a single band, the theory for the Hofstadter model predicts FQH states at the band fillings (\ref{eq:FillingFactorJainEquiv}) in bands of Chern number $C_H$ for the Hofstadter spectrum. As in the case of Chern number one bands \cite{app-Scaffidi:2012dx, app-Wu:2012ky, app-Wu:2013ii}, one can construct a sequence of many-body Hamiltonians that adiabatically deform the single-particle Hamiltonian of a suitable Hofstadter model to the corresponding Chern insulator model, as long as the Chern numbers for the relevant (occupied) bands are matched $C=C_H$. 

For the case of $C=1$, we see that the adiabatic connection to the problem of continuum Landau levels \cite{app-Scaffidi:2012dx, app-Wu:2013ii} can be constructed also in the Hofstadter model, as the deformation within the gap $C=1$, $D=0$ yields the continuum limit for $n_\phi\to 0$. The continuum limit is attractive as it provides a flat Berry curvature, which is believed to be optimal for the stability of Chern insulators (see e.g., \cite{app-Parameswaran:2012cu, app-2014PhRvB..90p5139R,app-Jackson:2014vj}). More generally, we can take effective continuum limits to perfectly flat $|C|>1$ bands at the points $n_\phi\to 1/|C|$. These points are analogous to the $n_\phi\to 0$ case in terms of their large MUC, and their perfectly flat dispersion and Berry curvature. Note that a perfectly flat Berry curvature can also be obtained for finite size systems of square geometry \cite{app-Scaffidi:2014gf}. 

In order to construct the adiabatic continuation of a $|C|>1$ Chern bands of a generic tight-binding model to a Hofstadter model at finite $\Delta n_\phi$, there are two proven strategies. Firstly, one may establish a mapping between these models at the level of the single-particle orbitals, either in the Wannier basis \cite{app-Barkeshli:2012kw,app-Scaffidi:2012dx,app-Wu:2012eq} or more favourably in the Bloch basis \cite{app-Wu:2013ii}.
Secondly, one may pursue a strategy of embedding the target model as a subset of a Hofstadter lattice and then adiabatically tuning the respective bonds \cite{app-Wu:2012ky}.

\section{Ground-State Degeneracy and Quasiparticle Properties}
\label{sec:quasiparticles}

For the sake of clarity, let us briefly re-state the results by Kol and Read in our notation. Following their work \cite{app-Kol:1993wv}, the Hall conductance $\sigma_{xy}$ and quasiparticle charge $e^*_\text{qh}$ follow simply from considering a Laughlin-type thought experiment of flux insertion for the composite fermions. The Hall conductivity $C^*$ of composite fermions (the $\sigma_{xy}^\text{mf}$ of Kol and Read) follows from the application of the Streda formula to the composite fermion system. To transport a composite fermion, one needs to insert not just one flux quantum, but one also needs to supply $kC^*$ additional flux to compensate for the gauge-flux attached to the particles in (\eref{eq:flux_attachment}). Again, from Laughlin's argument the quasiparticle charge also follows, leading to Kol and Read's results \cite{app-Kol:1993wv}
\be
\label{eq:SigmaHallAndQPCharge}
\sigma_{xy}= \frac{C^*}{1 + k C^*}\frac{e^2}{h},\quad e^*_\text{qh}= \frac{e}{1 + k C^*}.
\ee
Finally, their Chern-Simons theory yields the ground-state degeneracy $d_\text{GS}$ and statistics angle $\delta_\text{qh}$ for the (Abelian) statistics of the quasiparticles as
\be
\label{eq:GS_DegAndStatistics}
d_\text{GS}= |1 + k C^*|,\quad \delta_\text{qh}= \pi \left(1-\frac{2}{1 + k C^*}\right).
\ee
As an immediate consequence of these relations, we see that the many-body Chern number of the $d_\text{GS}$-fold degenerate ground-state manifold is equal to the Chern number of the composite fermion gap $C^*$. This matches the result obtained for a state with $|C^*|=2$ in our previous work \cite{app-2009PhRvL.103j5303M}.

\section{Particle-Entanglement Spectra}
\label{sec:PES}

For the $r=1$ candidates of our series (\ref{eq:FillingFactorJainEquiv}), the particle entanglement spectra yield the known entanglement sector count of states of the color-entangled Halperin states \cite{app-Sterdyniak:2013du}. Note also that the entanglement spectra for $C=1$ bands of the Harper-Hofstadter model have been studied in Ref.~\cite{app-Sterdyniak:2012jo}.

\begin{figure*}[t]
\begin{center}
\includegraphics[width=0.75\textwidth]{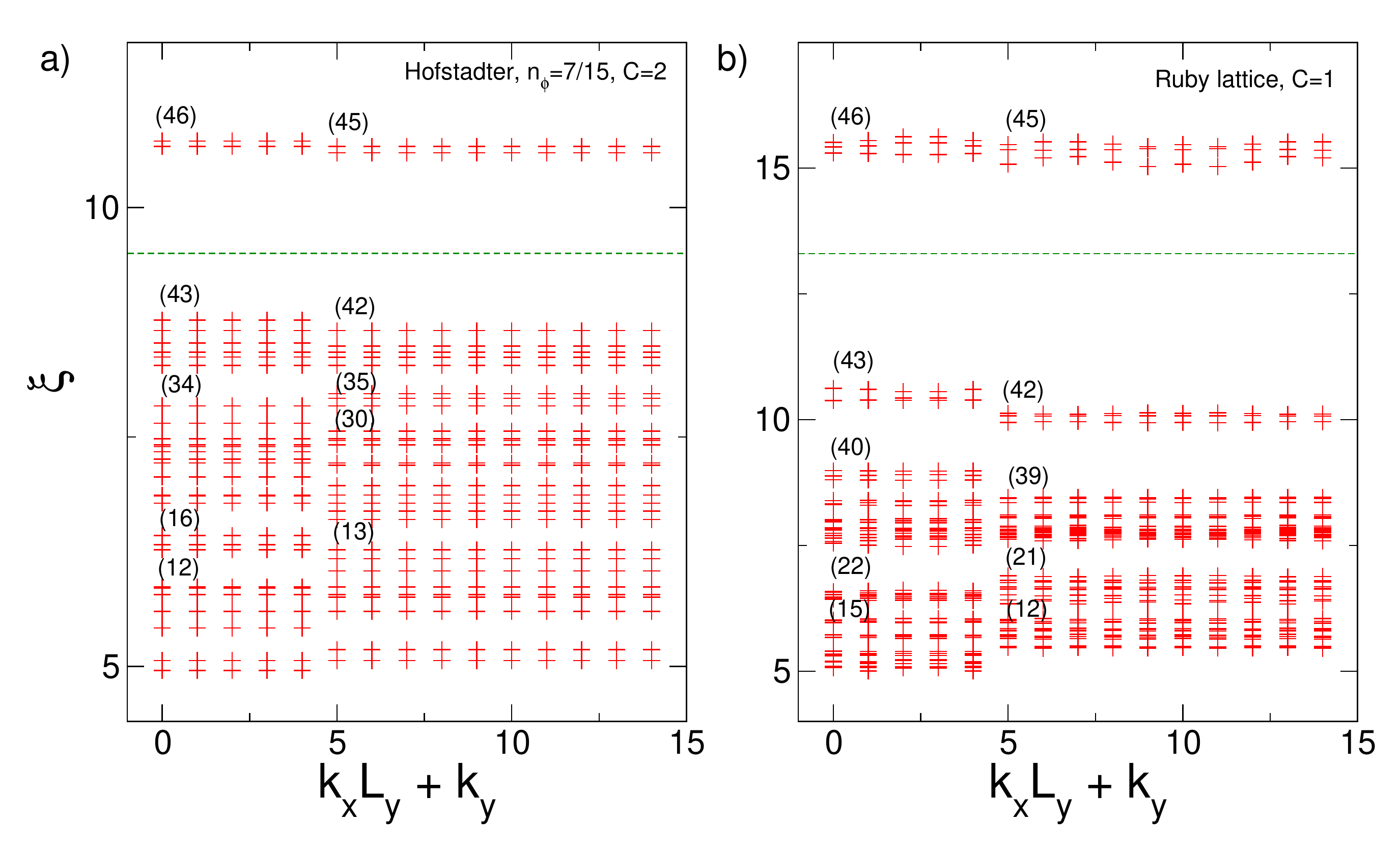}
\caption{Particle entanglement spectra for composite fermion / hierarchy states of bosons at $\nu=2/3$. Both panels show spectra at $N=10$ bosons, with $N_A=3$. We display the entanglement energy $\xi$ as a function of momentum sector $K=k_x L_y+k_y$. a) PES for for the $C=2$ band of the Harper-Hofstadter Hamiltonian at flux density $n_\phi=7/15$. b) Reference case of the PES in the $C=1$ band of the Ruby lattice model, at the parameters chosen by Liu \emph{et al.} \cite{app-2013PhRvB..87t5136L}. Note the different scales on the respective $y$-axes. Numbers in parentheses indicate the cumulative count of eigenstates from the bottom of the spectrum. The green dotted line indicates the principal entanglement gap.}
\label{fig:PES_nu_2_3}
\end{center}
\end{figure*}

The analysis of particle entanglement spectra for additional states with $r\neq 1$ for the series of hierarchy / composite fermion states (\ref{eq:FillingFactorJainEquiv}) is difficult, as no simple counting rule for these spectra is known even for Chern number $C=1$ bands. Correspondingly, these spectra have been little studied, and a detailed analysis of the particle entanglement spectra would be a complex topic on its own. We will not enter into details in the current paper, however, it is interesting to compare the structure of the PES at the same filling fractions in different Chern bands for the sake of illustration. In Fig.~\ref{fig:PES_nu_2_3}, we consider the case of the entanglement spectrum for $N_A=3$ out of $N=10$ bosons at filling factor $\nu=2/3$ as realized in the $C=2$ Hofstadter bands at flux density $n_\phi=7/15$ (a) in comparison to the PES in the $C=1$ of the Ruby lattice model (b). The latter was previously discussed and shown in Fig.~6 of Ref.~\cite{app-2013PhRvB..87t5136L}. 

We find both similarities and differences between the corresponding spectrum of bosons at $\nu=2/3$ in a $C=2$ band, Fig.~\ref{fig:PES_nu_2_3}a), and the $C=1$ data in Fig.~\ref{fig:PES_nu_2_3}b). The overall scale of the entanglement energies is smaller for $C=2$: we find the largest entanglement energies are $\xi^{C=2}_\text{max}\simeq 10.7$, compared to $\xi^{C=1}_\text{max}\simeq 15.5$ for $C=1$. Correspondingly, the entanglement gap is also smaller, at $\Delta^{C=2}_\xi \simeq 1.8$, compared to $\Delta^{C=1}_\xi \simeq 4.5$. However, the fraction of entanglement gap to the overall entanglement bandwidth $\delta_\xi=\Delta_\xi / (\xi_\text{max}-\xi_\text{min})$ are quite similar with $\delta^{C=2}_\xi \simeq 0.31$, compared to $\delta^{C=1}_\xi \simeq 0.42$. 

For the $C=1$ case, it was possible to infer the expected number of states from the corresponding continuum FQHE in the lowest Landau level by virtue of a FCI to FQHE mapping \cite{app-2013PhRvB..87t5136L}. By virtue of this property, that work found that the count of entanglement eigenvalues below the entanglement gap [indicated by a green line in our Fig.~\ref{fig:PES_nu_2_3}b)] is identical to the count inferred from the vanishing properties of the $\nu=2/3$ composite fermion state. In the case of higher Chern numbers, no corresponding continuum models are known, so we cannot take this correspondence as a given. For the example under consideration, we find that the number of entanglement states (as indicated in Fig.~\ref{fig:PES_nu_2_3}) below the principal gap is the same in both the $C=2$ and the $C=1$ models, with 43 of a total 46 states in sectors with $k_x=0$ and 42 of 45 states in sectors with $k_x\neq 0$. However, the blocking of states below that gap is clearly different, so the lower-lying $C=1$ entanglement gap identified with the Gaffnian state in \cite{app-2013PhRvB..87t5136L} does not carry over. Other differences appear in the dependency of the entanglement count on the number of particles $N_A$ that are traced out from the density matrix. For example, $N_A=4$, we find 200 states below the largest entanglement gap for the $C=1$ Ruby lattice model, while there are only 198 states for the $C=2$ Hofstadter model.

In conclusion, we find that the entanglement spectrum of the higher Chern number case bears a considerable resemblance to its $C=1$ counterpart. Hence, the data suggests that the physical nature of the ground state is also the same in both cases, i.e.~that of an incompressible quantum liquid. A detailed analysis would require the development of theoretical predictions for the entanglement sector count of the hierarchy wave functions, which could be based on, for example, the thin-torus framework of Ref.~\cite{app-2014PhRvB..89o5113W}.

\begin{figure}[t]
\begin{center}
\includegraphics[width=\columnwidth]{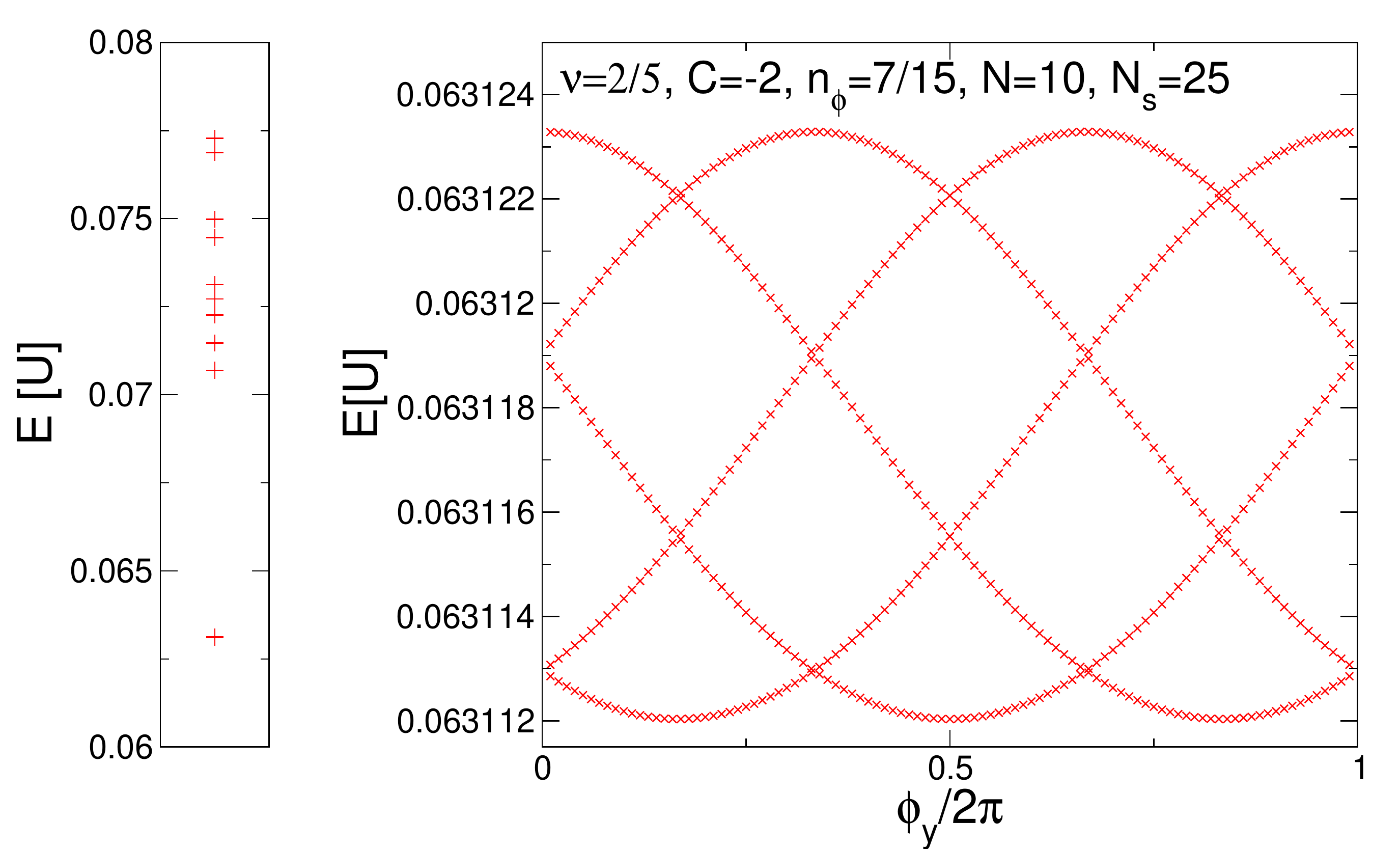}\\
\includegraphics[width=\columnwidth]{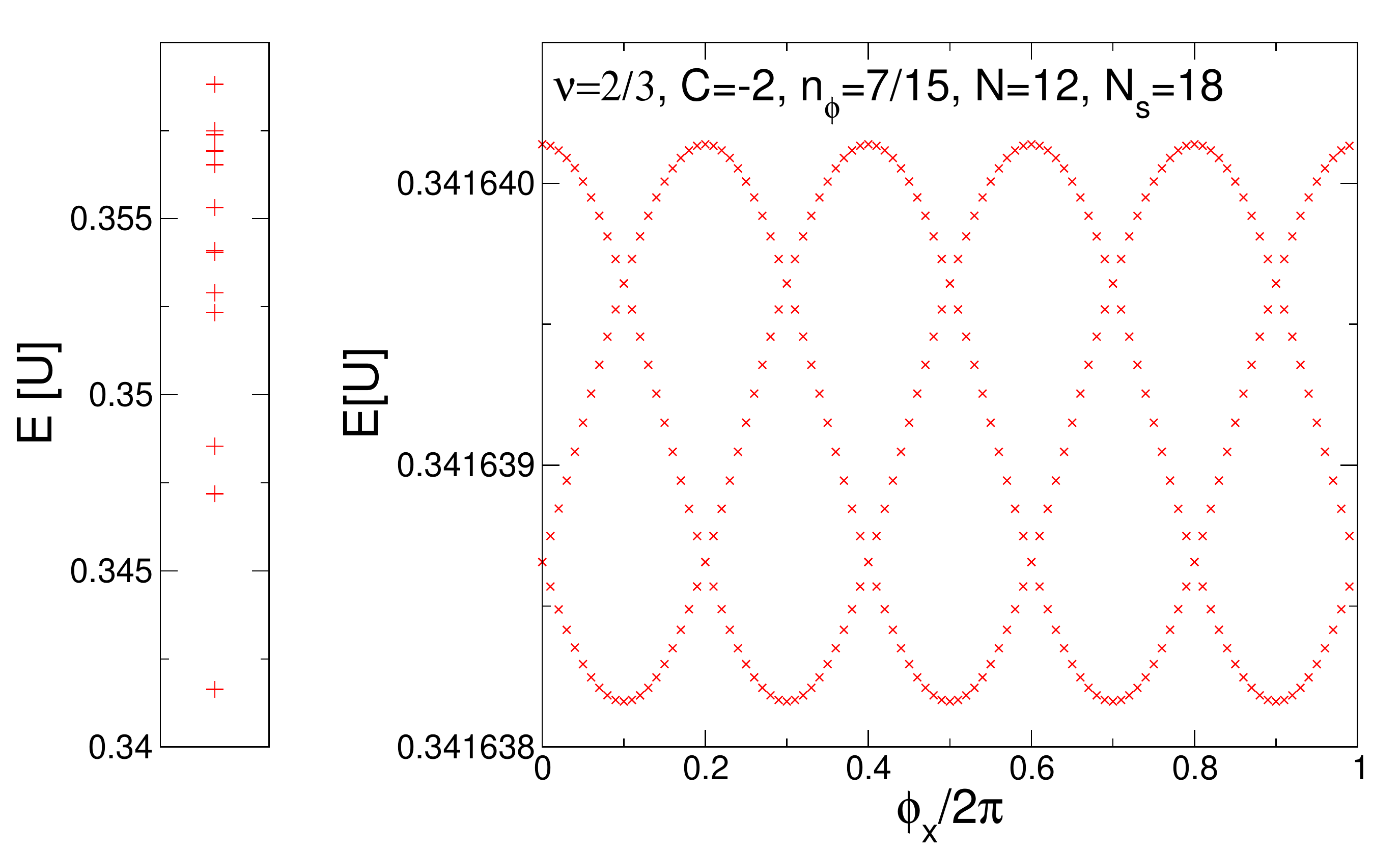}
\caption{Spectral flow for FCI states in a $|C|=2$ Hofstadter band at $n_\phi=7/15$. Note the splitting of the ground states is orders of magnitude smaller than the gap to the first excited state, so we choose to show the flow within the ground-state manifold only. For reference, the full spectrum at $\phi_x=\phi_y=0$ is shown to the left of the respective panels.  Top: Spectral flow for $N=10$ bosons in a $C=-2$ band at filling $\nu=2/5$. Bottom:  Spectral flow for $N=12$ bosons in the same $C=-2$ band at filling $\nu=2/3$. Geometries are further described in the main text.
 }
\label{fig:Flow}
\end{center}
\end{figure}

\section{Spectral Flow}
\label{sec:Flow}

The spectral response to the insertion of flux provides a characteristic feature of incompressible quantum liquids. In our numerical simulations, we can probe this response by applying twisted boundary conditions to our periodic simulation cell, notably by setting boundary conditions for single-particle particle states as $\psi(\br+L_\alpha \mathbf{\hat e}_\alpha)=e^{i \phi_\alpha} \psi(\br)$. For a state in the fractional quantum Hall regime, we expect to see spectral flow among the states of the ground-state manifold, but no mixing among the latter and any higher excited states. Indeed, this scenario is realized in the candidates for fractional quantum Hall or FCI states at filling factors of the series (\ref{eq:FillingFactorJainEquiv}). 

For illustration, we examine the spectral flow for states with $|r|=2$ in a $|C|=2$ band in Fig.~\ref{fig:Flow}, displaying filling factors $\nu=2/5$ and $\nu=2/3$ for positive and negative flux attachment, respectively. We consider the same geometries shown in Fig.~\ref{fig:Spectra}a,b) of the main text. In both these cases, all $d_\text{GS}$ ground states are found in the $(k_x,k_y)=(0,0)$ sector, so these states are not protected by symmetry from hybridizing with each other. 

For the $\nu=2/5$ state, we choose a system of $N=10$ particles $N_\text{site}=15\times 25$ sites, in a gauge with magnetic unit cell of $l_x=3$ times $l_y=5$ sites (and $N_s=25$ states in the lowest band). 
We find that the (near-) degenerate ground-states cycle among each other, remaining well separated in energy from all excited states under variation of $\phi_y$. 
Considering near-degenerate points of pairs of ground-states as crossings, the ensemble of states returns to its initial ordering and energies only after an insertion of five units of flux, owing to the five-fold degeneracy. 

For the state at $\nu=2/3$, we choose a system of $N=12$ particles in $N_\text{site}=18\times 15$ sites ($N_s=18$), using the same magnetic unit cell of $3\times 5$ sites. A similar picture ensues, demonstrating a flow among the ground state energies with a $3\times 2\pi$ periodicity of the spectrum in $\phi_x$.

Like the other features that we have examined, these results support our identification of the partially filled Hofstadter bands at the fillings (\ref{eq:FillingFactorJainEquiv}) as well-formed topological quantum liquids.

\end{document}